\documentstyle[12pt]{article}

\begin{document}
{\sf \begin{center} \noindent {\Large \bf Kinematic fast cosmic dynamos in non-inflationary phases of ellipsoidal universe}\\[3mm]

by \\[0.3cm]{\sl L.C. Garcia de Andrade}\\

\vspace{0.5cm} Departamento de F\'{\i}sica
Te\'orica -- IF -- Universidade do Estado do Rio de Janeiro-UERJ\\[-3mm]
Rua S\~ao Francisco Xavier, 524\\[-3mm]Cep 20550-003, Maracan\~a, Rio de Janeiro, RJ, Brasil\\[-3mm]
Electronic mail address: garcia@dft.if.uerj.br\\[-3mm]
\vspace{2cm} {\bf Abstract}
\end{center}
\paragraph*{}
Cosmic kinematic fast dynamo is found in non-inflationary phases of
an ellipsoidal anisotropic cosmological metric background solution
of Einstein field equations of general relativity. The magnetic
field is amplified inside the universe and spatially periodically. A
finite resistivity is assumed, and a nonsingular flow velocity is
aligned with the magnetic field which is orthogonal to a plane which
is analog to a galactic plane in astrophysics. Magnetic field
components is stretched along the z-direction and a cosmic dynamo is
created in the spirit of Zeldovich stretch, twist and fold (STF)
dynamo generation mechanism. In the inflationary phase of the planar
symmetric universe, the primordial magnetic field decays and the
galactic plane expands as a de Sitter $(2+1)-spacetime$ and the
eccentricity of the ellipsoidal universe, tends to vanish with
inflation. We may conclude that, as far as the present model is
concerned, anti-dynamos are obtained in inflationary
phases.\vspace{0.5cm} \noindent {\bf PACS numbers:}
\hfill\parbox[t]{13.5cm}{02.40.Hw-Riemannian geometries}

\newpage
 \section{Introduction}
 Cosmic and galactic dynamos have been investigated in recent
 year mainly by Brandenburg et al \cite{1} and Ruzmaikin, Shukurov and Sokoloff \cite{2}  which considered a turbulent
 magnetohydrodynamics (MHD) \cite{3} and mean field MHD previously investigated by Krause and R\"{a}dler \cite{3}. More recently
 cosmic dynamo solution of the Einstein field equation representing the Friedmann-Robertson-Walker (FRW) early
 universe cosmic dynamo have been presented by Brandenburg et al \cite{4}. They investigated in detail the numerical solution of
 hydromagnetic turbulence of primordial fields. These four dimensional pseudo-Riemannian space-time dynamos are
 four-dimensional extensions of Arnold et al \cite{5} kinematic cat dynamo based on Zeldovich \cite{6} STF dynamo mechanism. Though the authors considered that he magnetic
 fields generate local bulk motion, and that it is in large scales consistent with homogeneity and isotropy
 of the cosmological model, here we argue that the anisotropic models could be consider exactly if one drops the exigence of
 large scales. In this brief report we present a very simple non-turbulent analytical solution of
 kinematic dynamo MHD equations, which represents a cosmic dynamo embedded into an
 ellipsoidal universe recently discovered by Campanelli et al \cite{7} to explain the CMB quadrupole problem.
 To simplify matters we consider that the plasma flow velocity of
 the  highly conducting cosmological fluid is aligned with the
 magnetic field which is orthogonal to an analogous galactic plane,
 which greatly simplifies computations, since in this case the cross
 product term $\vec{u}{\times}\vec{B}$ vanishes. Magnetic flow speed
 fields are both dependent of small though finite resistivity of the
 plasma flow. Our solution presents a peculiarity that while of of
 the directions contracts the other remain constant in the
 plane-symmetric spacetime. Since the primordial magnetic field is
 proportionally to the small resistivity of the highly
 conductive cosmic flow, it is highly amplified in this
 ellipsoidal cosmology. Anisotropic cosmological models are also very suitable to deal with temperature
 anisotropies in the universe. The paper is organized as follows: Section II reviews the general relativistic (GR)
 plane-symmetric solution which to simplify matters we consider with just expansion or contraction of this
 universe. Section III presents the cosmic kinematic dynamo ellipsoidal universe solution in non-inflationary phases
 and snti-dynamos in de Sitter inflationary phase.
 Section IV presents the conclusions.
 \section{Plane-symmetric universes in Einstein gravity}
 The plane symmetric pseudo-Riemannian metric
 \begin{equation}
 ds^{2}=-dt^{2}+a^{2}(t)[dx^{2}+dy^{2}]+b^{2}(t)dz^{2}\label{1}
 \end{equation}
 Substitution of this metric into the Einstein equations yields the
 following differential equations
 \begin{equation}
 (\frac{\dot{a}}{a})^{2}+2\frac{\dot{a}}{a}\frac{\dot{b}}{b}=8{\pi}G{\rho}\label{2}
 \end{equation}
 \begin{equation}
 \frac{\ddot{a}}{a}+\frac{\ddot{b}}{b}+\frac{\dot{a}}{a}\frac{\dot{b}}{b}=-8{\pi}Gp_{||}\label{3}
 \end{equation}
\begin{equation}
 2\frac{\ddot{a}}{a}+(\frac{\dot{a}}{a})^{2}=-8{\pi}Gp_{\perp}
 \label{4}
 \end{equation}
where the anisotropic energy-stress tensor is given by
\begin{equation}
{T^{\mu}}_{\nu}=(\rho,-p_{||},-p_{\perp},-p_{\perp}) \label{5}
\end{equation}
and the Einstein equations are used in the form
\begin{equation}
{R^{\mu}}_{\nu}-\frac{1}{2}{{\delta}^{\mu}}_{\nu}R=8{\pi}G{T^{\mu}}_{\nu}
\label{6}
\end{equation}
where the indices are given as $(\mu=0,1,2,3)$ and R and
${R^{\mu}}_{\nu}$ are respectively the Ricci scalar and tensor. Here
we shall be concerned with a particular ellipsoidal solution where
$b(t)=b_{0}=const$ which reduce the Einstein field equations above
to
\begin{equation}
 (\frac{\dot{a}}{a})^{2}=8{\pi}G{\rho}\label{7}
 \end{equation}
 \begin{equation}
 \frac{\ddot{a}}{a}=-8{\pi}Gp_{||}\label{8}
 \end{equation}
\begin{equation}
 2\frac{\ddot{a}}{a}+(\frac{\dot{a}}{a})^{2}=-8{\pi}Gp_{\perp}
 \label{9}
 \end{equation}
 \section{Cosmic dynamos in non-inflationary phase}
 The MHD dynamo equations
\begin{equation}
{\nabla}.\vec{B}=\frac{1}{\sqrt{g}}{\partial}_{i}[{\sqrt{g}}g^{ij}B_{j}]=0\label{10}
\end{equation}
where $g_{ij}$ is the Riemannian $3D$ spatial part of the planar
symmetric metric. Here $\sqrt{g}=a^{2}b_{0}$ and $g$ is the
determinant of the 3D metric. The remaining
\begin{equation}
{\partial}_{t}\vec{B}+(\vec{u}.{\nabla})\vec{B}-(\vec{B}.{\nabla})\vec{u}={\eta}{\nabla}^{2}\vec{B}\label{11}
\end{equation}
where ${\eta}$  the resistivity. Since the second and third terms on
the LHS represent ${\nabla}{\times}(\vec{u}{\times}\vec{B})$ it
vanishes and the self induction equation (\ref{11}) reduces to
\begin{equation}
{\partial}_{t}\vec{B}={\eta}{\nabla}^{2}\vec{B}\label{12}
\end{equation}
To solve this equation let us assume that
\begin{equation}
\vec{B}=c(x,y){a}^{-2}\vec{k}\label{13}
\end{equation}
which fulfills equation (\ref{10}). Substitution of expression
(\ref{13}) into (\ref{12}) yields the following relation
\begin{equation}
\frac{b_{0}}{c}[{\partial}_{x}+{\partial}_{y}]c(x,y)=-\frac{2}{\eta}\frac{\dot{a}}{a}\label{14}
\end{equation}
Since each side of PDE depends on independent variables t, and
$(x,y)$ one can equate both sides to the same constant $c_{1}$ which
yields the two PDE
\begin{equation}
\frac{2}{\eta}\frac{\dot{a}}{a}=-c_{1}\label{15}
\end{equation}
and
\begin{equation}
\frac{b_{0}}{c}[{\partial}_{x}+{\partial}_{y}]c(x,y)=c_{1}\label{16}
\end{equation}
Solution of the first equation yields
\begin{equation}
a=e^{-\frac{c_{1}{\eta}}{2}t}\label{17}
\end{equation}
Note that if $c_{1}<0$ the galactic plane expands as a de Sitter
inflationary phase of a anisotropic universe. However, the
primordial magnetic field (\ref{13}) decays and a cosmic dynamo is
not obtained. On the other hand if the constant obeys the constraint
$c_{1}>0$ deflationary or contracting phase of the universe is
obtained and a primordial field of a fast cosmic kinematic dynamo
yields
\begin{equation}
\vec{B}=c(x,y)e^{pt}\vec{k}\label{18}
\end{equation}
where $p(\eta)=-\frac{c_{1}{\eta}}{2}>0$ yields the magnetic field
\begin{equation}
\vec{B}=exp{(pt)}c\label{19}
\end{equation}
where $p(\eta)=c_{1}{\eta}$ , which yields the fast cosmic dynamo
\cite{8}. To complete the solution we simply solve the PDE
\begin{equation}
\frac{b_{0}}{c}[{\partial}_{x}+{\partial}_{y}]c(x,y)=c_{1}\label{20}
\end{equation}
A simple particular solution of this equation yields
$c(x,y)=e^{\frac{ic_{1}}{b_{0}}(x+y)}$ which shows that the
primordial cosmic magnetic dynamo field is given by
\begin{equation}
\vec{B}=B_{0}e^{pt+\frac{ic_{1}}{b_{0}}(x+y)}\vec{k}\label{21}
\end{equation}
Now going back to dynamo flow we can compute it by solving the
equation
\begin{equation}
{\nabla}.\vec{u}=-\frac{1}{\rho}\frac{{\partial}{\rho}}{{\partial}t}\label{22}
\end{equation}
To solve this equation we need first to go back to the Einstein
field equations to obtain
\begin{equation}
{\rho}=\frac{p^{2}(\eta)}{8{\pi}G}\label{23}
\end{equation}
\begin{equation}
{p}_{||}=-{p^{2}(\eta)}{8{\pi}G}\label{24}
\end{equation}
\begin{equation}
{p}_{\perp}=-{\frac{3}{4}p^{2}(\eta)}{8{\pi}G}\label{25}
\end{equation}
which from expression (\ref{22}) the dynamo flow is incompressible
\begin{equation}
{\nabla}.\vec{u}=0\label{26}
\end{equation}
and from this expression we finally obtain the dynamo flow as
\begin{equation}
{\partial}_{z}(a^{2}{b_{0}}^{-1}{u}_{z})=0\label{27}
\end{equation}
\begin{equation}
{u}_{z}=F(x,y)e^{pt}+u_{0}
\end{equation}
where $u_{0}$ is an integration constant of the dynamo
incompressible flow. Note that when the resistivity vanishes the
dynamo flow is constant. Finally the eccentricity of the ellipsoidal
universe may be computed as
\begin{equation}
e=\sqrt{1-{b_{0}}e^{pt}}\label{28}
\end{equation}
and we note that in this universe the cosmic dynamo and resistivity
of the conductive fluid affects the eccentricity of the
universe.
\section{Conclusions} A more detailed account of the decay of MHD 3D magnetic
fields in the early turbulent universe can be found in the work of
Hindmarsh, Christensson and Brandenburg \cite{9}. However in this
brief report we addressed the problem of the existence of a cosmic
kinematic dynamo embbeded into an ellipsoidal universe. It is shown
that as far as this model is concerned inflationary phases do not
admit dynamos but only anti-dynamos. On the other hand
non-inflationary or deflationary phases allows for the presence of
cosmic fast dynamos. The eccentricity of the universe depends also
on the resistivity of the highly conducting fluid. In future
investigation of the primordial magnetic field we also consider
turbulent anisotropic universes as more realistic cosmic dynamo
models.
 \section*{Acknowledgements}
 Thanks are due to K. H. Tsui and A. Yassuda for helpful discussions on the subject of this paper and to CNPq and UERJ for financial supports.

\end{document}